\documentclass[12pt]{article}
\usepackage{graphicx}

\newcommand\pubnumber{ATL-PHYS-PROC-2018-117}
\newcommand\pubdate{\today}

\def\uppsala{Department of Physics and Astronomy\\
Uppsala University, SE-75120 Uppsala, SWEDEN}

\def\Title#1{\begin{center} {\Large #1 } \end{center}}
\def\Author#1{\begin{center}{ \sc #1} \end{center}}
\def\Address#1{\begin{center}{ \it #1} \end{center}}

\newcommand\pubblock{\rightline{\begin{tabular}{l} \pubnumber\\
         \pubdate  \end{tabular}}}
\newenvironment{Abstract}{\begin{quotation}  }{\end{quotation}}
\newenvironment{Presented}{\begin{quotation} \begin{center} 
             PRESENTED AT\end{center}\bigskip 
      \begin{center}\begin{large}}{\end{large}\end{center} \end{quotation}}

\def\beq{\begin{equation}}
\def\eeq#1{\label{#1}\end{equation}}
\def\eeqn{\end{equation}}

\def\beqa{\begin{eqnarray}}
\def\eeqa#1{\label{#1}\end{eqnarray}}
\def\eeqan{\end{eqnarray}}

\let\bar=\overbar

\def\Dslash{\not{\hbox{\kern-4pt $D$}}}
\def\dslash{\not{\hbox{\kern-2pt $\del$}}}

\def\msb{{\bar{\ssstyle M \kern -1pt S}}}

\usepackage{cite}
\usepackage{lineno}

\begin{document}
\begin{titlepage}
\pubblock

\vfill
\Title{Searches for Higgs boson pair production with ATLAS}
\vfill
\Author{Arnaud Ferrari, on behalf of the ATLAS Collaboration}
\Address{\uppsala}
\vfill
\begin{Abstract}
The search for Higgs boson pair ($HH$) production is at the core of the ATLAS experimental program, as it probes the 
Brout-Englert-Higgs mechanism, as well as new physics beyond the Standard Model. Based on the proton-proton collision 
data recorded by the ATLAS experiment at CERN's Large Hadron Collider in 2015 and 2016, three public results on the search 
for $HH \to bbbb$, $HH \to bb\gamma\gamma$ and $HH \to WW\gamma\gamma$ were recently released and are summarised 
in these proceedings.
\end{Abstract}
\vfill
\begin{Presented}
Thirteenth Conference on the Intersections of\\ Particle and Nuclear Physics (CIPANP 2018)\\~\\
Indian Wells, California, USA, May 29 to June 3, 2018
\end{Presented}
\vfill
\end{titlepage}
\def\thefootnote{\fnsymbol{footnote}}
\setcounter{footnote}{0}

\section{Introduction}

The discovery of the Higgs boson ($H$) by the ATLAS and CMS Collaborations at the Large Hadron Collider (LHC) in 
2012~\cite{Aad:2012tfa,Chatrchyan:2012xdj} has confirmed the Brout-Englert-Higgs (BEH) mechanism of electroweak 
symmetry breaking and mass generation~\cite{Englert:1964et,Higgs:1964pj}. The Higgs boson mass is about 
125~GeV~\cite{Aad:2015zhl} and its measured properties are in agreement with the Standard Model (SM) 
predictions~\cite{Khachatryan:2016vau}. However, the BEH mechanism does not only predict the existence 
of a massive scalar particle, but it also requires the Higgs boson to couple to itself. In order to measure the 
Higgs potential and thereby have a complete description of the SM, it is necessary to observe the production 
of Higgs boson pairs and measure the Higgs self-coupling $\lambda_{HHH}$. Also, any deviation from the 
SM predictions would open a window on new physics.\\

\begin{figure}[htbp]
  \centering
\includegraphics[width=0.32\textwidth]{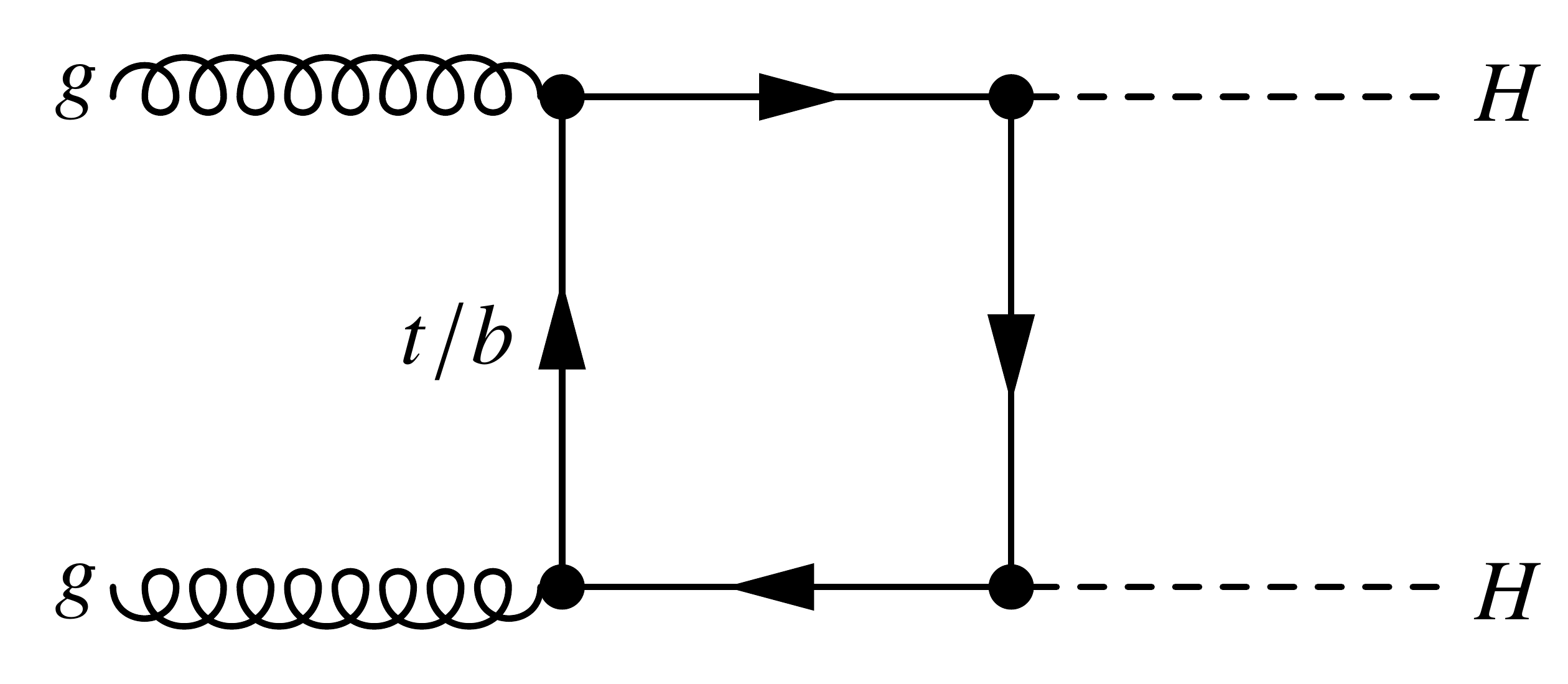}
\includegraphics[width=0.32\textwidth]{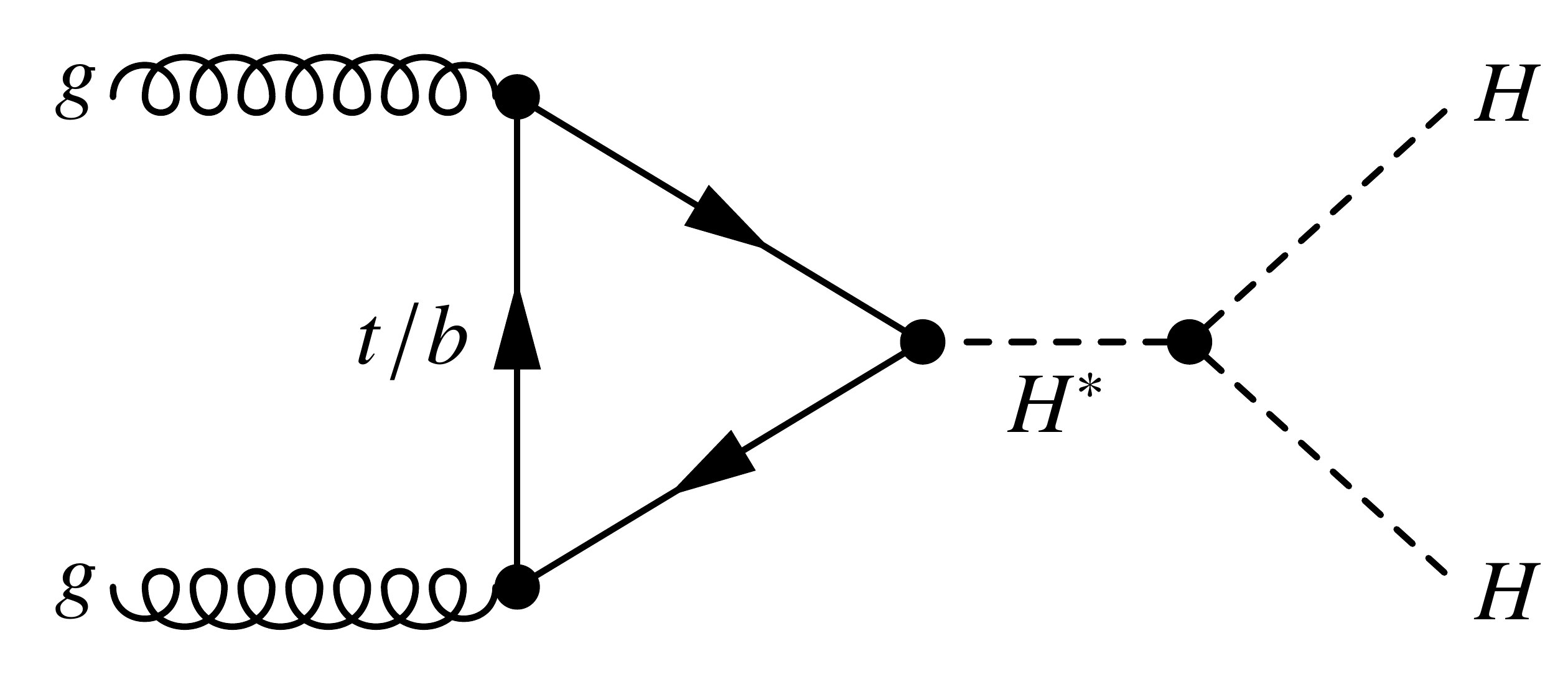}
\includegraphics[width=0.32\textwidth]{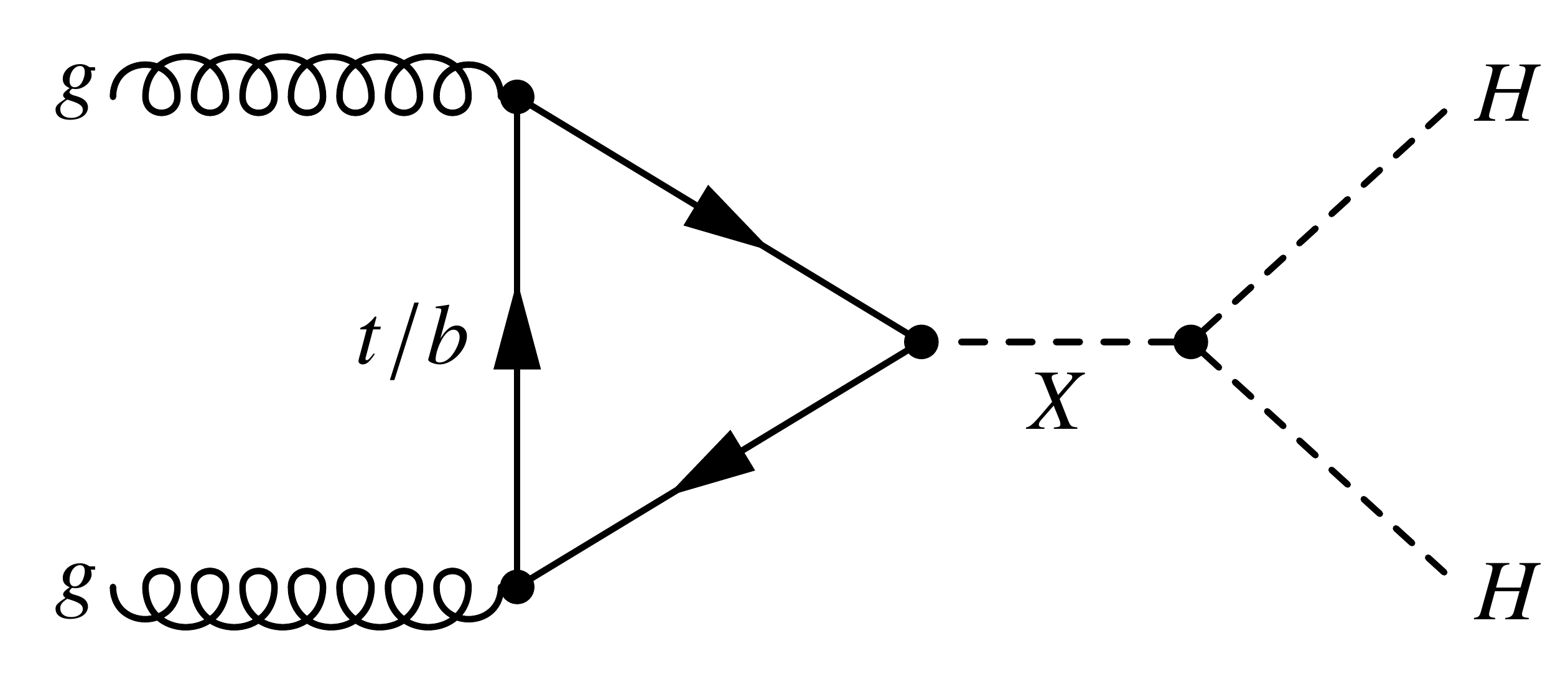}
\caption{\label{fig:hh-diagrams} Leading-order Feynman diagrams for $HH$ production in the gluon-gluon fusion mode, via 
(left) a heavy-quark loop and (centre) Higgs self-coupling in the SM, or through (right) an intermediate resonance ($X$).}
\end{figure}

In the SM, Higgs boson pairs ($HH$) can be produced either in a heavy-quark loop or via Higgs self-coupling,
see the left-hand and central diagrams of Figure~\ref{fig:hh-diagrams}, respectively. However, due to the destructive 
interference between the two diagrams, the SM Higgs boson pair production is only 33.41~fb at 13~TeV, as computed 
at the next-to-leading-order (NLO) in QCD and fully accounting for top-quark mass effects~\cite{deFlorian:2016spz,Grazzini:2018bsd}.
Still, the $HH$ cross-section can be significantly enhanced in several Beyond-the-Standard-Model (BSM) scenarios, e.g.\ through 
anomalous couplings, new contact interactions, or resonant production of $HH$ pairs, as illustrated in the right-hand diagram of 
Figure~\ref{fig:hh-diagrams}. 
Several BSM theories indeed predict the existence of heavy particles decaying into a pair of Higgs bosons. For instance, 
two-Higgs-doublet models (2HDMs)~\cite{Djouadi:2005gj,Branco:2011iw} have a second CP-even Higgs boson, which may be 
heavy enough to decay into two SM-like lighter Higgs bosons. Alternatively, $HH$ pairs can be produced in the decay of 
a spin-2 graviton, as predicted in the Randall-Sundrum model of warped extra dimensions~\cite{Randall:1999ee}.

Searches for non-resonant and resonant $HH$ production were conducted by the ATLAS collaboration, based on 
proton-proton ($pp$) collision data corresponding to an integrated luminosity of up to 36.1~fb$^{-1}$. The ATLAS 
experiment~\cite{PERF-2007-01,Capeans:2010jnh,ATLAS:2012rar} is a multi-purpose detector 
with a forward-backward symmetric cylindrical geometry and nearly $4\pi$ coverage in solid angle.\footnote{ATLAS uses a right-handed 
coordinate system with its origin at the nominal interaction point (IP) in the centre of the detector and the $z$-axis along the beam 
pipe. The $x$-axis points from the IP to the centre of the LHC ring, and the $y$-axis points upwards. Cylindrical coordinates 
$(r,\phi)$ are used in the transverse plane, $\phi$ being the azimuthal angle around the $z$-axis. The pseudo-rapidity is 
defined in terms of the polar angle $\theta$ as $\eta = -\ln \tan(\theta/2)$, and the angular distance is measured in units of 
$\Delta R \equiv \sqrt{(\Delta\eta)^{2} + (\Delta\phi)^{2}}$.} It consists of an inner tracking detector 
surrounded by a thin superconducting solenoid providing a 2~T axial magnetic field, electromagnetic and hadron calorimeters, 
and a muon spectrometer. A two-level trigger~\cite{Aaboud:2016leb,ATL-DAQ-PUB-2017-001} reduces the event rate to a 
maximum of 1~kHz prior to offline data storage. Three search results were recently published by the ATLAS Collaboration, in the 
$HH \to bbbb$~\cite{Aaboud:2018knk}, $HH \to bb\gamma\gamma$~\cite{Aaboud:2018ftw} and 
$HH \to WW\gamma\gamma$~\cite{Aaboud:2018ewm} channels. They are summarised in the following sections of these proceedings.

\section{Search for $HH \to bbbb$ in ATLAS}

With a branching fraction of 33\%, $HH \to bbbb$ has the largest event rate among all Higgs boson pair production channels. 
However, it also suffers from a large multi-jet background, which requires the use of novel data-driven estimation techniques. 
Depending on the $HH$ production mode and the probed mass range, two event topologies are considered: 
\begin{itemize}
\item For both the non-resonant $HH$ production mode and the search for a resonance decaying via $HH \to bbbb$ with a mass in the 
range 260--1400~GeV, a resolved topology is considered, where at least four anti-$k_T$ jets~\cite{Cacciari:2008gp} with a parameter 
radius $R=0.4$ are reconstructed with a transverse momentum $p_{\mathrm{T}} > 40~\mbox{GeV}$, and the number of $b$-tagged 
jets (i.e.\ jets compatible with a $b$-hadron decay) is required to be exactly four (the efficiency of the working point is 70\% in simulated 
$t\bar{t}$ events). Such events are recorded based on a combination of $b$-tagged jet triggers. Due to an inefficiency of vertex 
reconstruction and thereby $b$-tagging at the trigger level, a fraction of the data collected in 2016 was not retained, hence the 
integrated luminosity for this event topology is only 27.5~fb$^{-1}$.
\item For the search for a resonance decaying via $HH \to bbbb$ with a mass in the range 800--3000~GeV, a boosted topology 
is considered, where at least two anti-$k_T$ jets with a larger parameter radius $R=1.0$ are reconstructed (including one firing the 
corresponding trigger). The leading (sub-leading) large-$R$ jet is required to have $p_{\mathrm{T}} > 450~(250)~\mbox{GeV}$. 
Then, $b$-tagging is performed on track-jets with a parameter radius $R=0.2$ and at least one $b$-tag per large-$R$ jet is 
required, which yields three event categories with two, three or four $b$-tags.
\end{itemize}

\subsection{Resolved topology}

\begin{figure}[htbp]
\centering
\includegraphics[height=0.2\textheight]{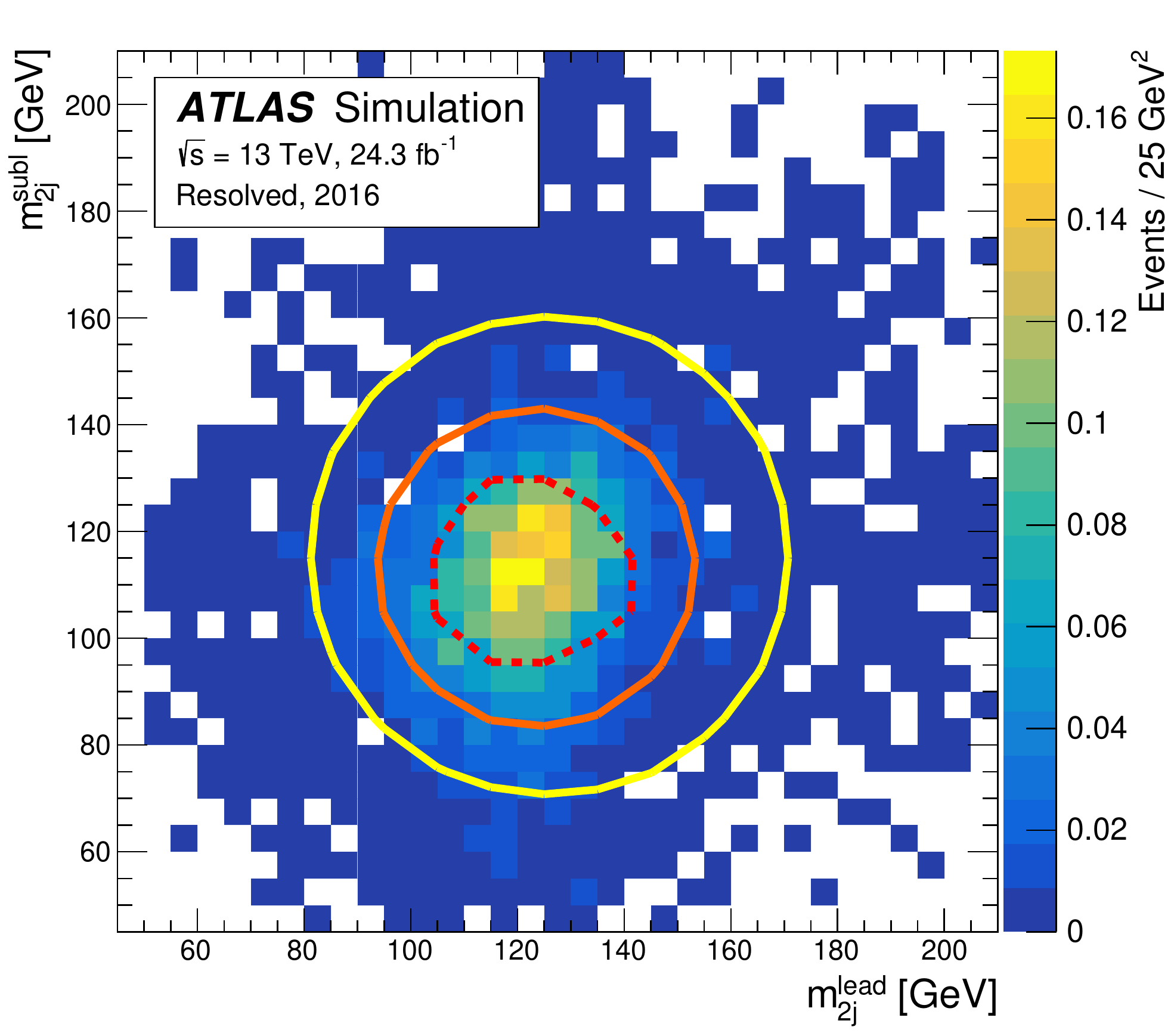}
\includegraphics[height=0.2\textheight]{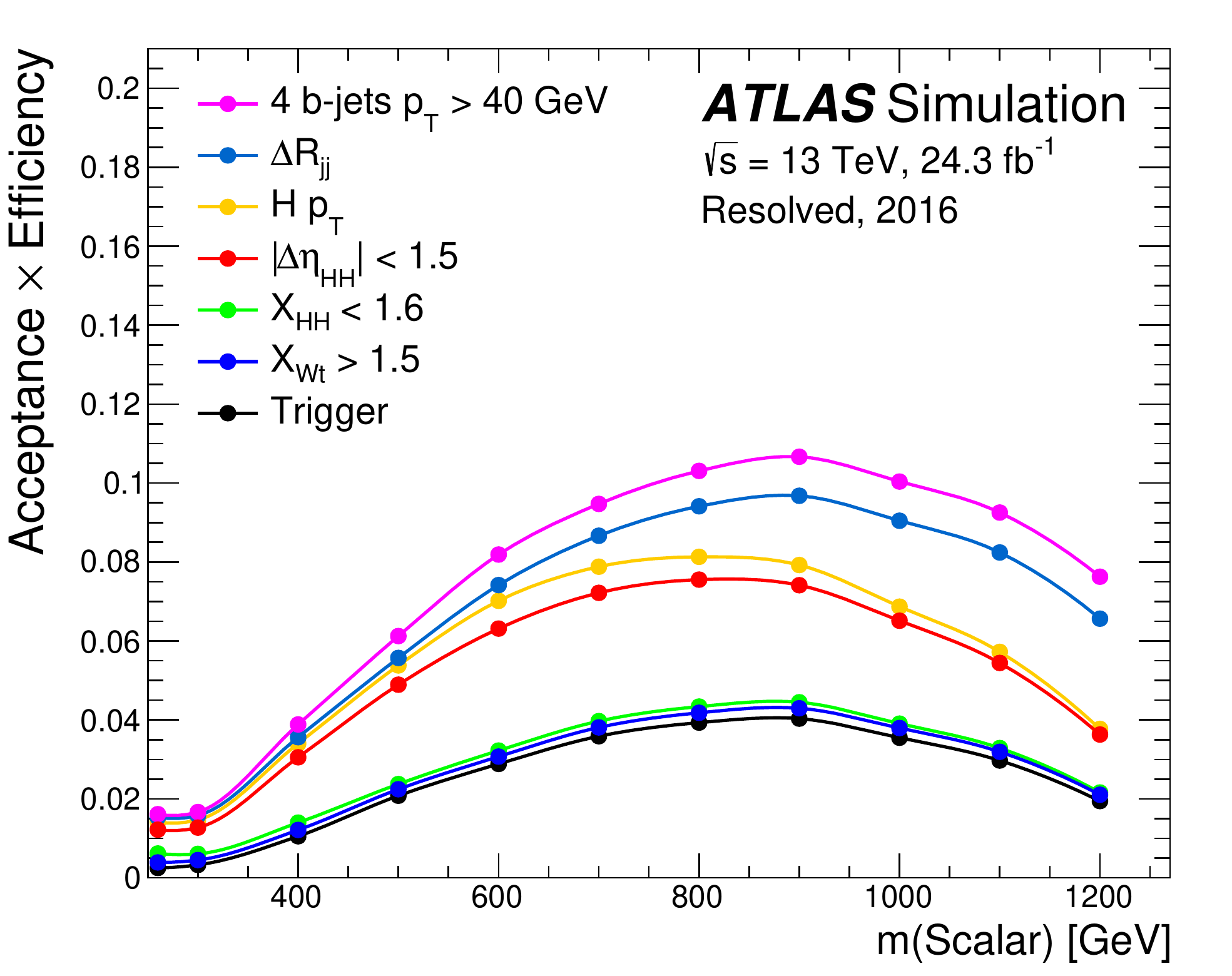}
\includegraphics[height=0.2\textheight]{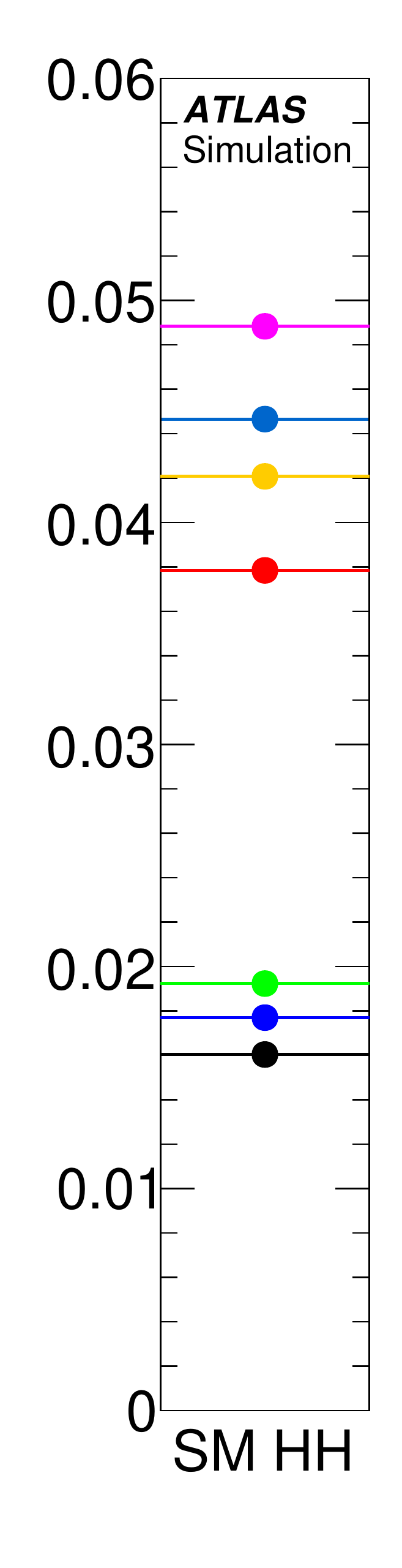}
\caption{\label{fig:bbbb-first} Analysis regions in the ($m_{2j}^{\mathrm{lead}}$;\,$m_{2j}^{\mathrm{sub-lead}}$) phase-space 
(left) where the area within the red dashed line is the signal region, while the area between the orange and yellow full lines is the 
sideband region used for data-driven background estimation, and event selection efficiencies for a spin-0 resonance decaying to 
a Higgs boson pair (centre) and non-resonant $HH$ production (right)~\cite{Aaboud:2018knk}.}
\end{figure}

For events with a resolved topology, the four jets with highest $b$-tagging scores are used. The selection and pairing of jets 
into Higgs boson candidates is performed using angular distances between jets ($\Delta R_{jj}$), the four-jet invariant mass 
$m_{4j}$ and differences in $m_{2j}$. Then, $m_{4j}$- and $m_{2j}$-dependent requirements on the $p_{\mathrm{T}}$ and mass 
of the Higgs boson candidates are applied. Events in which a three-jet combination is compatible with a top-quark decay are 
vetoed to reduce the $t\bar{t}$ background contamination. As a result, the signal region is defined by a small area in the 
($m_{2j}^{\mathrm{lead}}$;\,$m_{2j}^{\mathrm{sub-lead}}$) phase-space centered at (120~GeV;\,110~GeV), corresponding to 
the reconstructed masses of the two (leading and sub-leading) $H \to bb$ candidates, as illustrated by the red dashed line in 
the left-hand plot of Figure~\ref{fig:bbbb-first}. The efficiency of the various event selections is shown in the central (right-hand) 
plot of Figure~\ref{fig:bbbb-first} for various masses of a spin-0 resonance decaying to a Higgs boson pair (non-resonant $HH$ 
production).

The dominant multi-jet background is estimated with data. A data sample is built with the nominal event selection, 
but the number of $b$-jets is requested to be exactly two: one $H$ candidate is reconstructed from the two $b$-tagged jets, and 
the other one from two non-$b$-tagged jets. Re-weighting is then applied to this $2b$+$2j$ sample. The weights are derived 
by comparing $2b$+$2j$ and $4b$ samples in a sideband region, located between the orange and yellow full lines in the 
($m_{2j}^{\mathrm{lead}}$;\,$m_{2j}^{\mathrm{sub-lead}}$) phase-space: a per-non-$b$-tagged-jet factor is obtained by 
comparing jet multiplicities and a global event weight is obtained from the ratio of $4b$ and $2b$+$2j$ templates (after 
subtracting $t\bar{t}$ events) for five variables sensitive to differences in $b$-tagging. As for the $t\bar{t}$ background, 
its shape is taken from simulation, but the normalisations of the fully-hadronic and semi-leptonic $t\bar{t}$ backgrounds are 
determined together with the multi-jet event yield by a simultaneous fit in three background-enriched regions of the sideband 
region. A validation of the background estimate is performed in a dedicated region, which is between the orange full 
line and the red dashed line in the left-hand plot of Figure~\ref{fig:bbbb-first}, and a good agreement between the 
predicted and measured $m_{4j}$ distribution is found.\\

\begin{figure}[htbp]
\centering
\includegraphics[height=0.2\textheight]{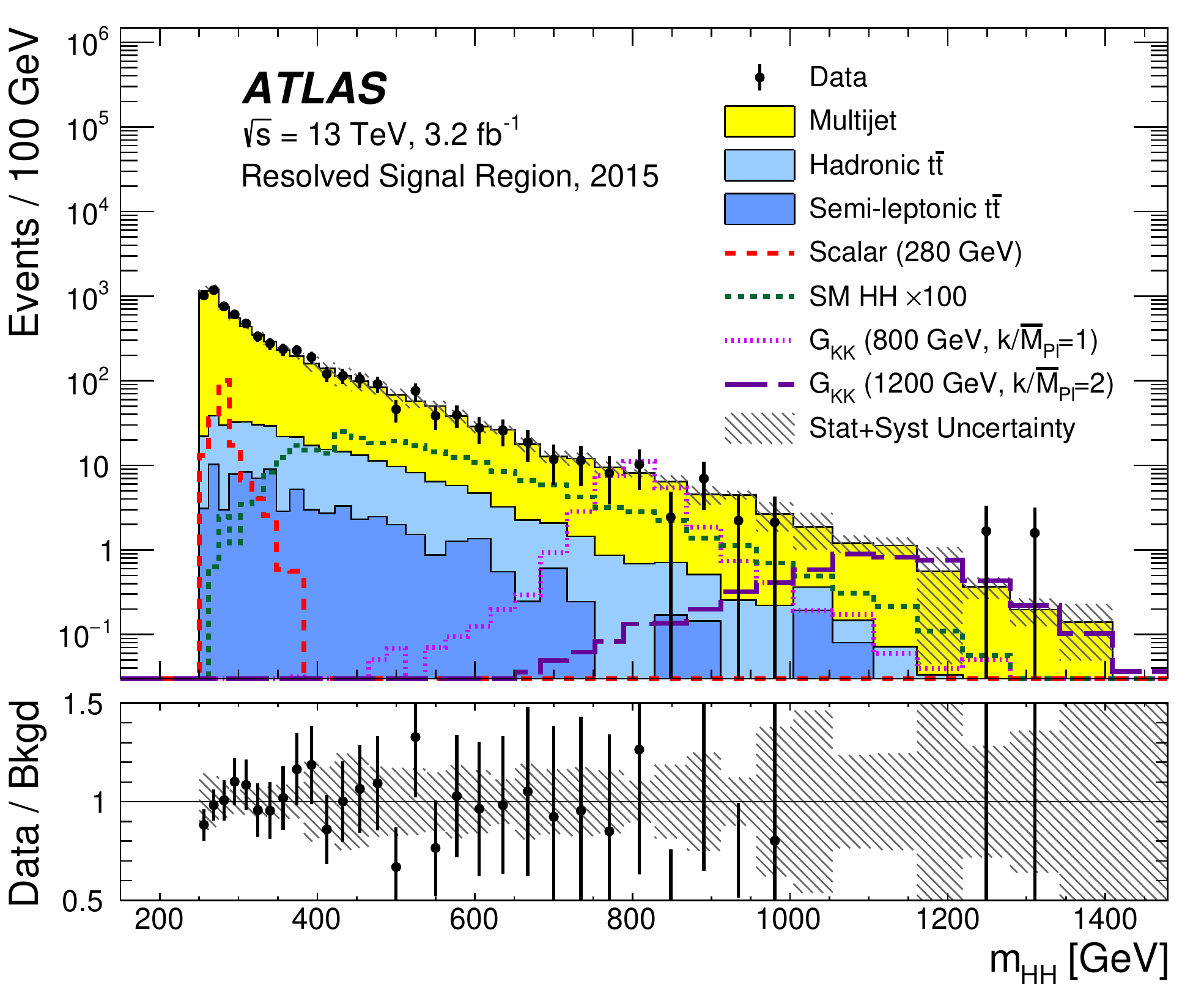}
\includegraphics[height=0.2\textheight]{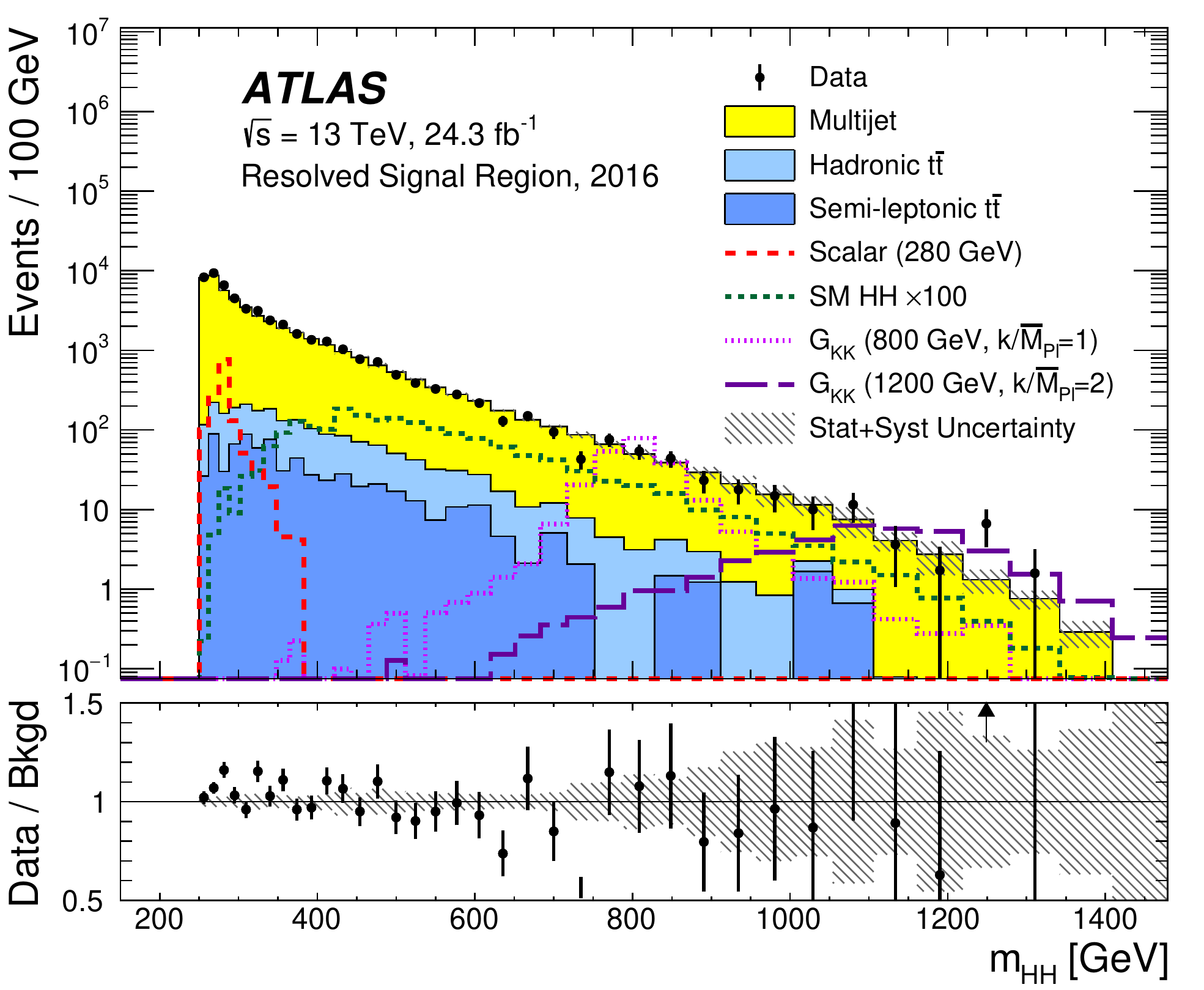}
\caption{\label{fig:bbbb-second} Reconstructed invariant mass of the two $H \to bb$ candidates in the signal region of the resolved 
topology, split by data-taking year~\cite{Aaboud:2018knk}.}
\end{figure}

Figure~\ref{fig:bbbb-second} shows the invariant mass distribution the two Higgs boson candidates in the signal 
region, split between the 2015 and 2016 datasets, as different trigger configurations were used. The largest local 
deviation has a statistical significance of 3.6$\sigma$ at 280~GeV, while the global significance is 2.3$\sigma$. 

\subsection{Boosted topology}

\begin{figure}[htbp]
\centering
\includegraphics[height=0.16\textheight]{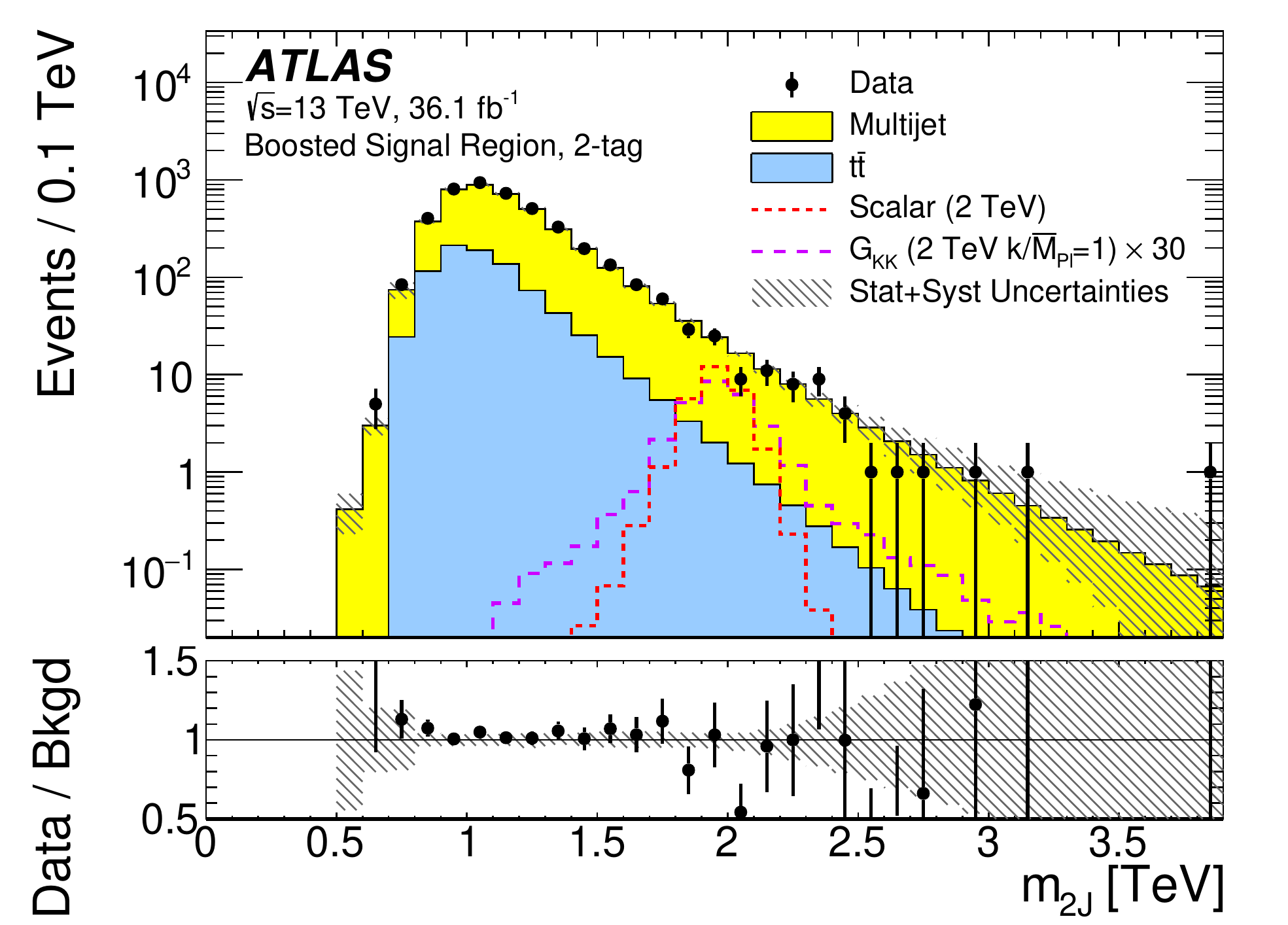}
\includegraphics[height=0.16\textheight]{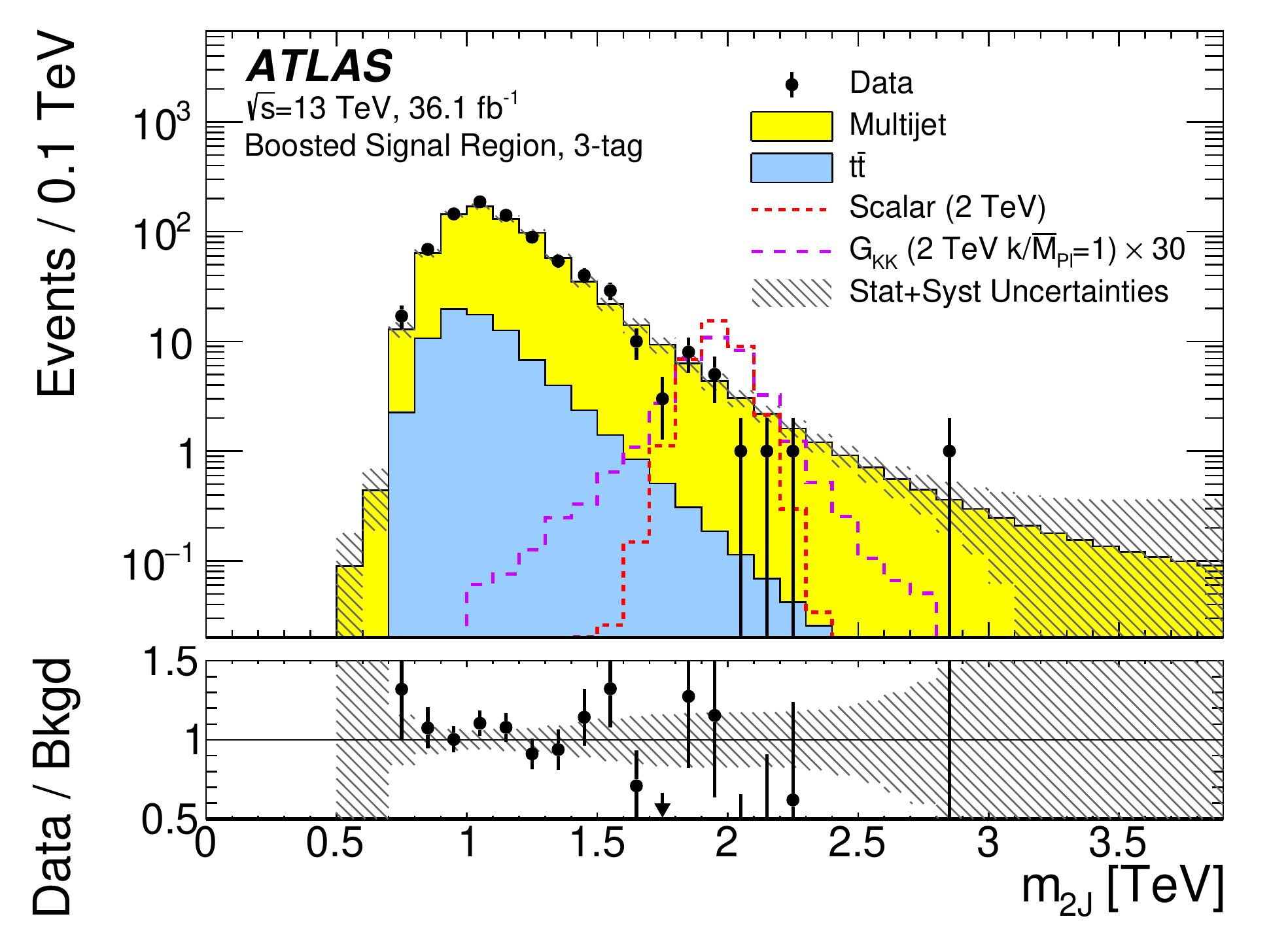}
\includegraphics[height=0.16\textheight]{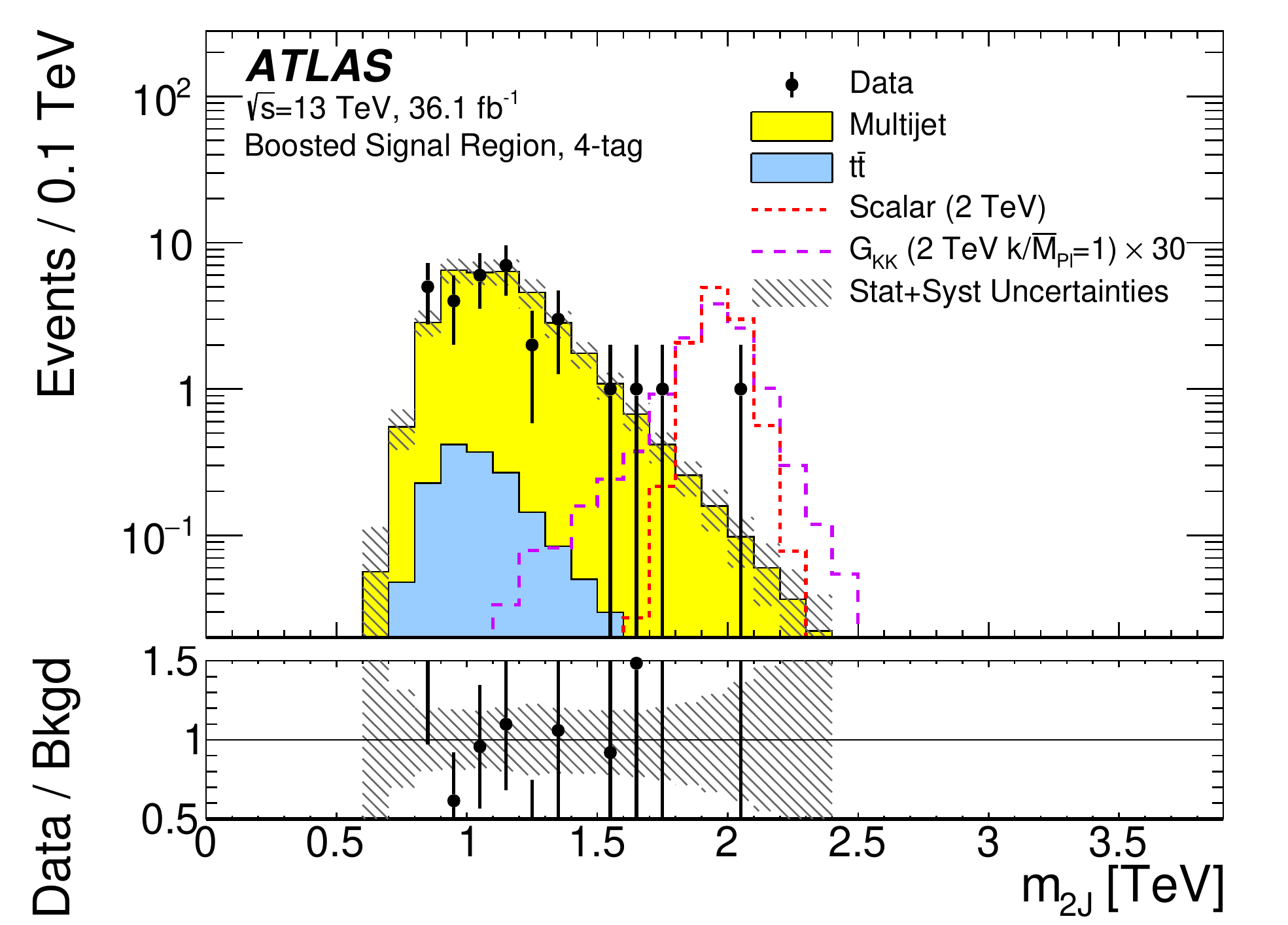}
\caption{\label{fig:bbbb-third} Reconstructed invariant mass of the two $H \to bb$ candidates in the signal region of the boosted 
topology, split by number of $b$-tags~\cite{Aaboud:2018knk}.}
\end{figure}

In the boosted topology, the two large-$R$ jets with highest $p_{\mathrm{T}}$ are used in order to reconstruct the two 
$H \to bb$ candidates, with an angular distance $|\Delta \eta_{JJ}| < 1.7$. Following further requirements on the jet 
masses, a signal region is defined in the ($m_{J}^{\mathrm{lead}}$;\,$m_{J}^{\mathrm{sub-lead}}$) phase-space centered at 
(124~GeV;\,115~GeV), corresponding to the reconstructed masses of the two (leading and sub-leading) $H \to bb$ candidates, 
each reconstructed as a single large-$R$ jet. Similarly to the data-driven estimation method employed in the case of resolved 
topologies, multi-jet templates are built from "lower-tagged" event selections (i.e. in which one of the large-$R$ jet has no 
$b$-tagged track-jet and at least one failing $b$-tagging) and the kinematic distributions of the non-$b$-tagged $J$ is 
re-weighted in order to mimic a $H$ candidate. The shape of the $t\bar{t}$ background is taken from simulation. Finally, 
the normalisation of the backgrounds is obtained from binned likelihood fits of the leading large-$R$ jet mass distribution 
in a sideband region. After successful validation of the background model in a dedicated region between the sideband and 
signal regions of the ($m_{J}^{\mathrm{lead}}$;\,$m_{J}^{\mathrm{sub-lead}}$) phase-space, the $m_{JJ}$ distribution 
in the signal region is used as a discriminant, after correction of the large-$R$ jet momenta by $m_H/m_J$, separately 
for events with two, three and four $b$-tags, see Figure~\ref{fig:bbbb-third}. 

\subsection{Results}

\begin{figure}[htbp]
\centering
\includegraphics[height=0.24\textheight]{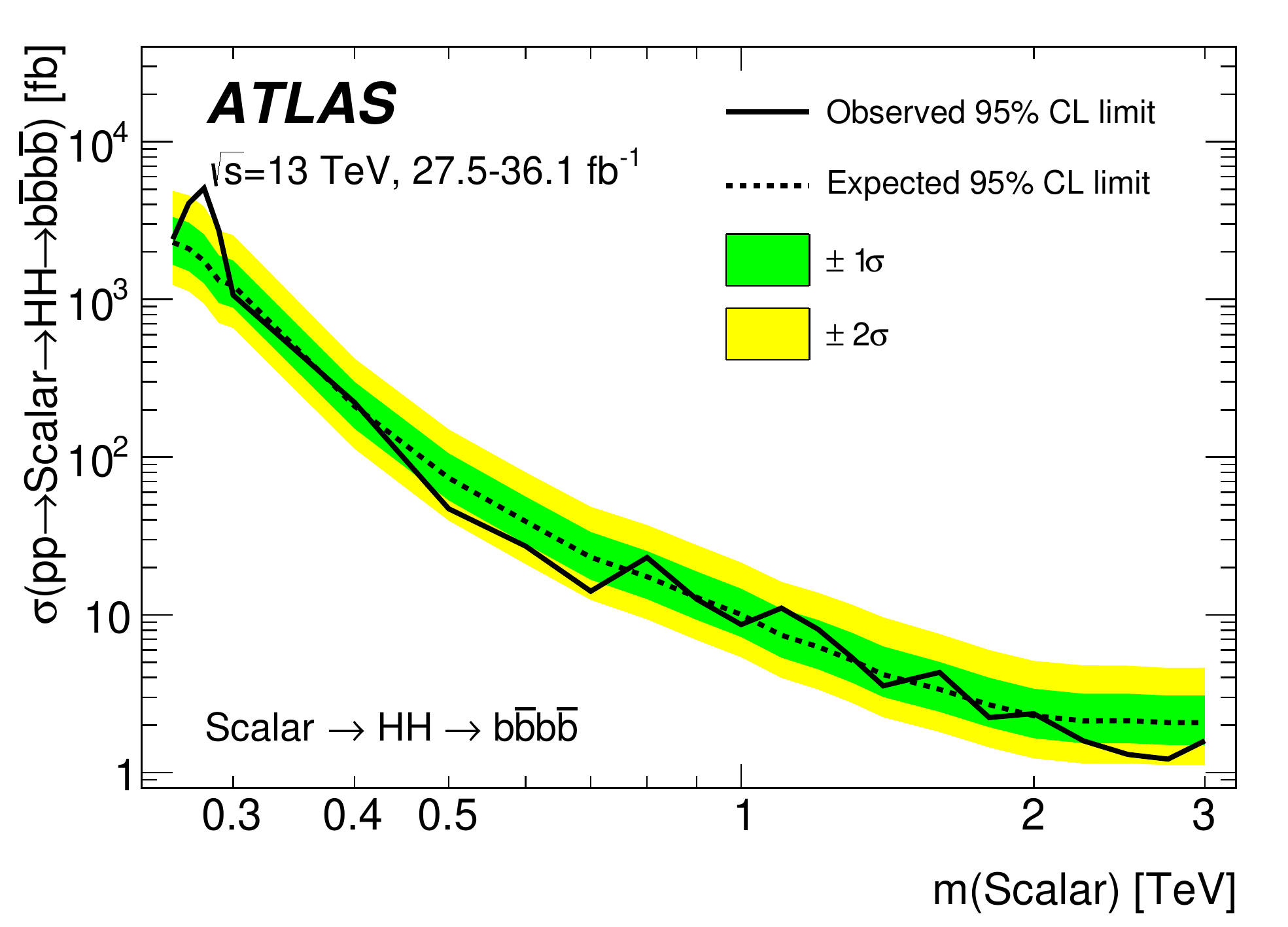}
\caption{\label{fig:bbbb-fourth} 95\% CL limits on Higgs boson pair production from a spin-0 resonance, as a function of the hypothetised 
mass, from the search for $HH \to bbbb$ in ATLAS~\cite{Aaboud:2018knk}.}
\end{figure}

The 95\% confidence-level (CL) exclusion limits on the non-resonant $HH$ production are shown in Table~\ref{tab:bbbb}, in 
units of the SM prediction. In the case of $HH \to bbbb$ production through a spin-0 resonance, a statistical combination of 
the resolved and boosted topologies is performed in the mass range where they overlap, i.e.\ 800--1400~GeV. The 95\% CL 
upper limits on the resonant production cross-section are shown in Figure~\ref{fig:bbbb-fourth}.
 
\begin{table}[htbp]
\begin{center}
\caption{95\% CL limits on non-resonant Higgs boson pair production, from the search for $HH \to bbbb$ in ATLAS.}
\vspace*{2mm}
\begin{tabular}{cccccc} 
\hline 
Observed &  $-2\sigma$ & $-1\sigma$ & Expected &   $+1\sigma$ & $+2\sigma$ \\
\hline
13.0 &  11.1 & 14.9 & 20.7 & 30.0 & 43.5 \\
\hline
\end{tabular}
\label{tab:bbbb}
\end{center}
\end{table}

\section{Search for $HH \to bb\gamma\gamma$ in ATLAS}

The search for $HH \to bb\gamma\gamma$ suffers from a low signal branching fraction (0.26\%), but the presence of a di-photon system 
with a good mass resolution provides a very clean signature. The event selection starts with a di-photon trigger, with transverse energy 
($E_{\mathrm{T}}$) thresholds at 25 and 35~GeV. Two photons with an invariant mass $m_{\gamma\gamma}$ between 105 and 
160~GeV are selected, with $E_{\mathrm{T}}/m_{\gamma\gamma}$ above 0.35 (0.25) for the leading (sub-leading) photon. 
The event selection proceeds with requesting at least two central jets with $p_{\mathrm{T}} > 25~\mbox{GeV}$, vetoing 
events with more than two $b$-tags. The signal region consists of events with exactly two $b$-tags (with the same working 
point as in the search for $HH \to bbbb$,) and events with one $b$-tag, however with a tighter working point corresponding 
to a 60\% efficiency. In that latter case, a Boosted Decision Tree (BDT) is used to assign a second jet to the $H \to bb$ candidate. 
In addition to the requirements on the photons and $b$-jets, two sets of event selections are applied:
\begin{itemize}
\item loose selection, used in the search for non-resonant production with a Higgs self-coupling different from the SM prediction and 
in the search for a spin-0 resonance of mass 260--500~GeV decaying via $HH \to bb\gamma\gamma$: the $p_{\mathrm{T}}$ of the 
leading (sub-leading) jet is greater than 40~(25)~GeV, the invariant mass of the two jets is $80~\mbox{GeV} < m_{jj} < 140~\mbox{GeV}$ 
and, in the case of the search for resonant $HH \to bb\gamma\gamma$ production, $m_{\gamma\gamma}$ must be within 4.7~GeV of 
$m_H$;
\item tight selection, used in the search for SM non-resonant production and in the search for a spin-0 resonance heavier than 
500~GeV decaying via $HH \to bb\gamma\gamma$: the $p_{\mathrm{T}}$ of the leading (sub-leading) jet is greater than 
100~(30)~GeV, the invariant mass of the two jets is $90~\mbox{GeV} < m_{jj} < 140~\mbox{GeV}$ and, in the case of the search 
for resonant $HH \to bb\gamma\gamma$ production, $m_{\gamma\gamma}$ must be within 4.3~GeV of $m_H$.
\end{itemize}

\subsection{Search for non-resonant $HH \to bb\gamma\gamma$ production}

\begin{figure}[htbp]
\centering
\includegraphics[height=0.21\textheight]{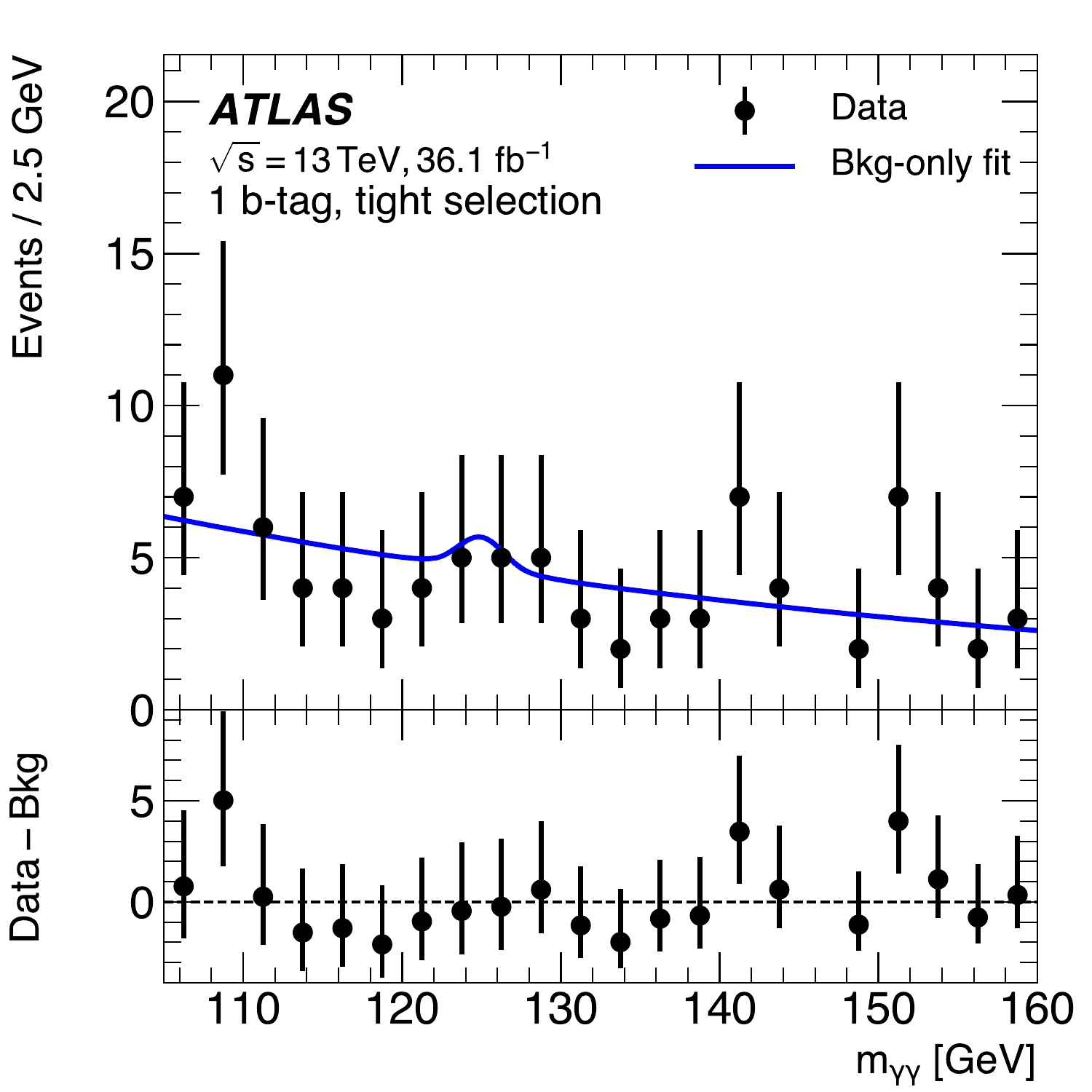}
\includegraphics[height=0.21\textheight]{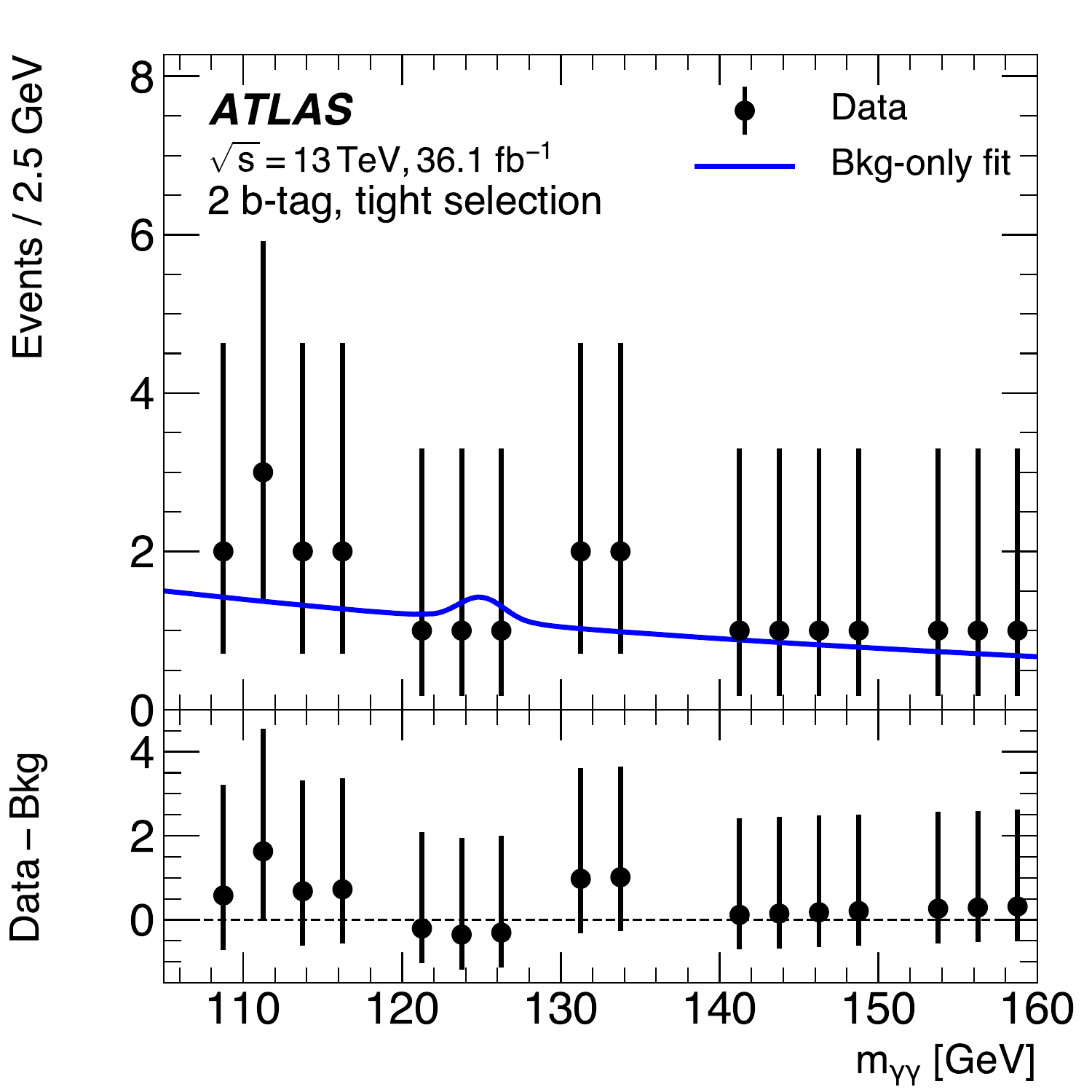}
\caption{\label{fig:bbyy-first} Di-photon mass spectrum in signal regions of the search for  $HH \to bb\gamma\gamma$ in ATLAS, 
with one (left) or two (right) $b$-tags, after applying a tight event selection~\cite{Aaboud:2018ftw}.}
\end{figure}

The analysis strategy used in the search for non-resonant $HH \to bb\gamma\gamma$ production is to extract the signal from 
the $m_{\gamma\gamma}$ distribution. Both the $HH$ signal and the single-$H$ backgrounds are taken from simulation, and 
$m_{\gamma\gamma}$ is parameterised with a double-sided Crystal-Ball function. On the other hand, the continuum background 
of multi-jet and multi-photon events is modelled by a fit to the data, with a first-order exponential function. This fit function was 
found to minimise the spurious signal, i.e.\ the bias measured by fitting a signal+background model to a background-only sample. 
Figure~\ref{fig:bbyy-first} shows the $m_{\gamma\gamma}$ distribution in the signal region with one and two $b$-tags, obtained 
with the tight selection. In the absence of a statistically significant excess with respect to the SM prediction, 95\% CL exclusion 
limits are set of the Higgs boson pair production cross-section, in units of the SM prediction, see Table~\ref{tab:bbyy}.

\begin{table}[htbp]
\begin{center}
\caption{95\% CL limits on non-resonant Higgs boson pair production, from the search for $HH \to bb\gamma\gamma$ in ATLAS.}
\vspace*{2mm}
\begin{tabular}{cccc}
\hline
Observed & $-1\sigma$ & Expected & $+1\sigma$  \\
\hline
22 & 20 & 28 & 40 \\
\hline
\end{tabular}
\label{tab:bbyy}
\end{center}
\end{table}

\begin{figure}[htbp]
\centering
\includegraphics[height=0.21\textheight]{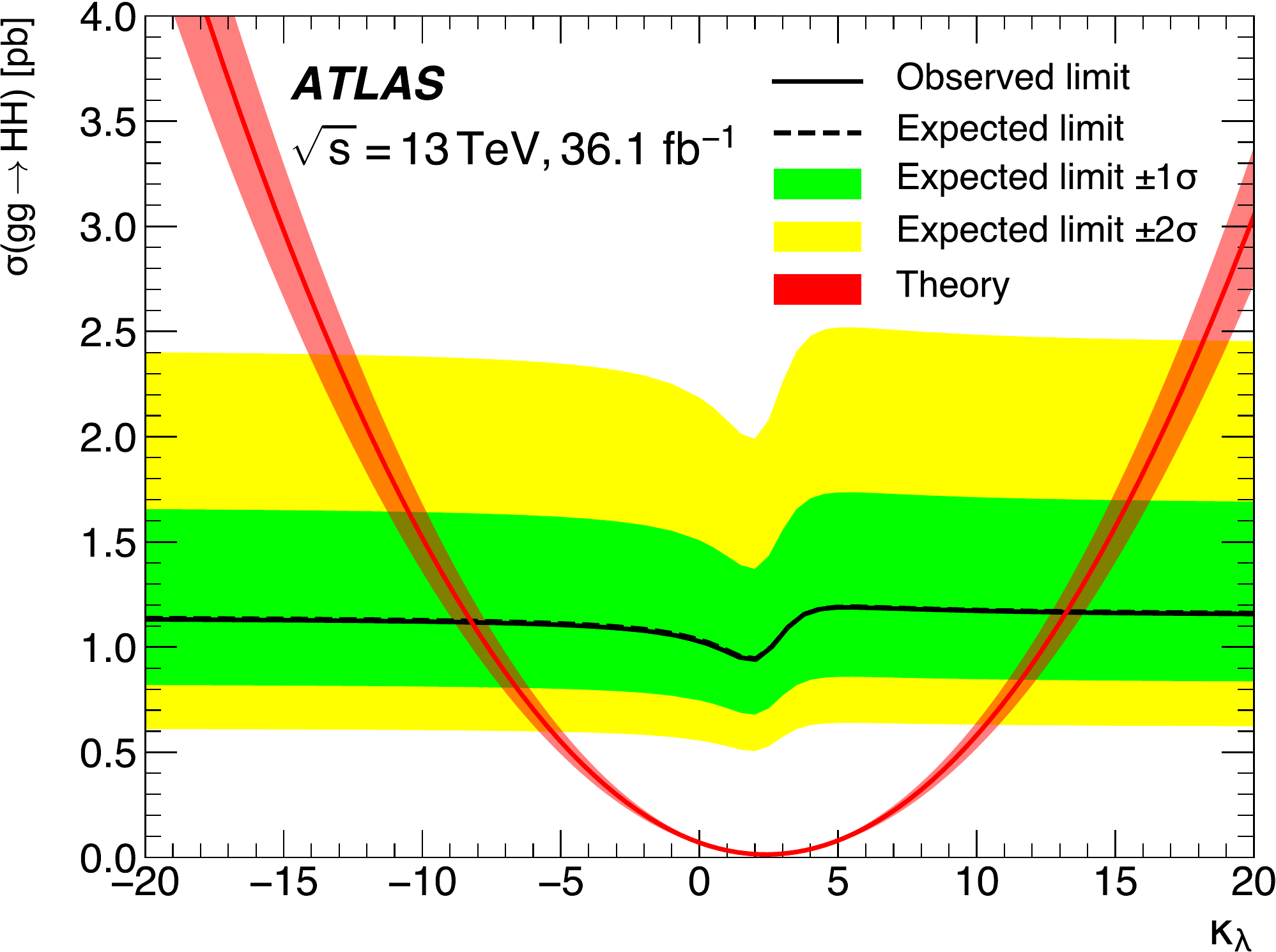}
\caption{\label{fig:bbyy-second} 95\% CL limits on non-resonant Higgs boson pair production as a function of $\kappa_{\lambda}$, 
from the search for $HH \to bb\gamma\gamma$ in ATLAS. The red line indicates the theoretical prediction with its 
uncertainty~\cite{Aaboud:2018ftw}.}
\end{figure}

Higgs self-coupling values that differ from the SM prediction affect the signal acceptance. The variations of the exclusion 
limits with $\kappa_{\lambda}$, defined as the ratio of the Higgs self-coupling $\lambda$ to its predicted value in the SM, 
$\lambda_{\mathrm{SM}}$, are computed with the loose event selection and are shown in Figure~\ref{fig:bbyy-second}. 
As a result, $\kappa_{\lambda}$ is constrained at the 95\% CL to be between $-8.2$ and 13.2.

\subsection{Search for resonant $HH \to bb\gamma\gamma$ production}

\begin{figure}[htbp]
\centering
\includegraphics[height=0.21\textheight]{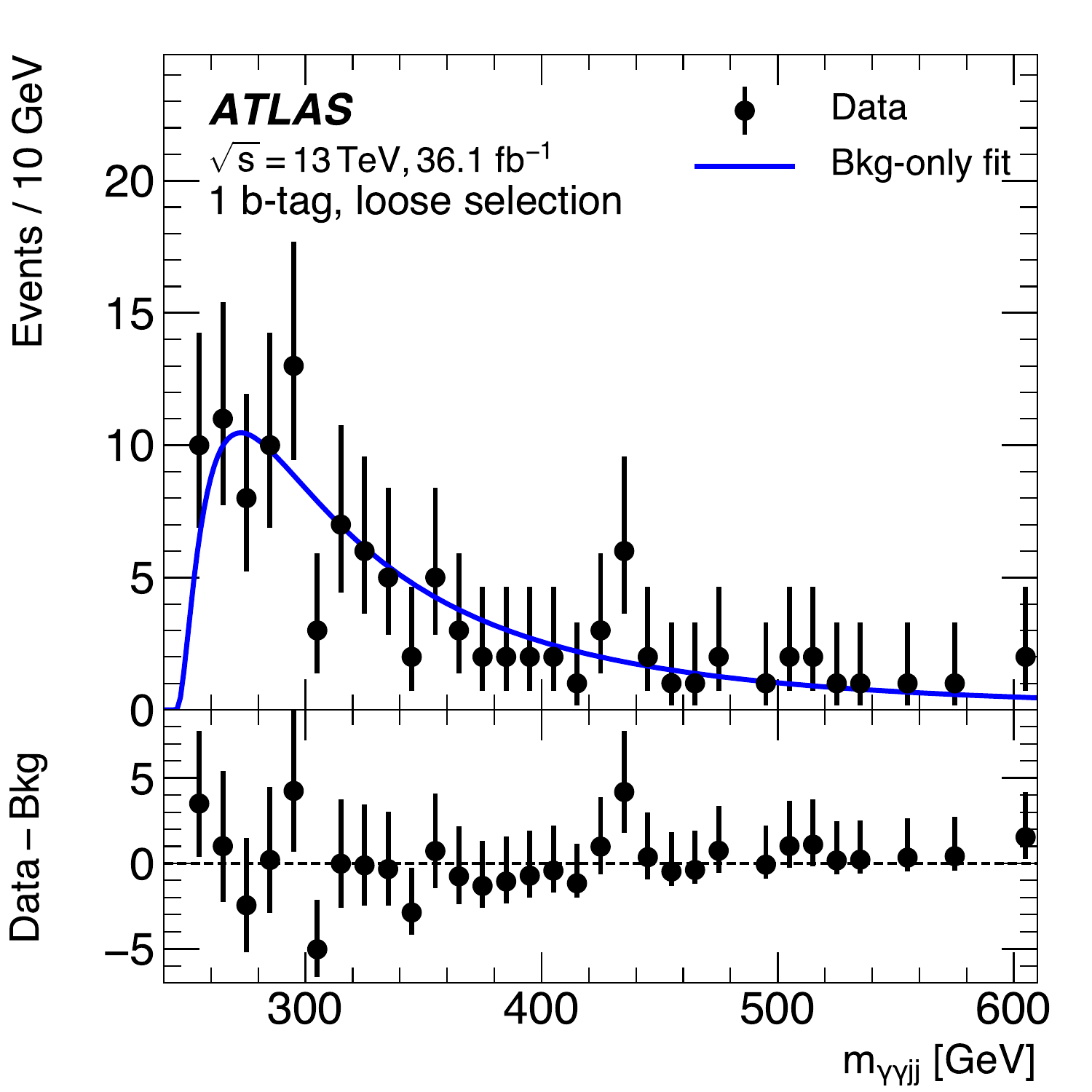}
\includegraphics[height=0.21\textheight]{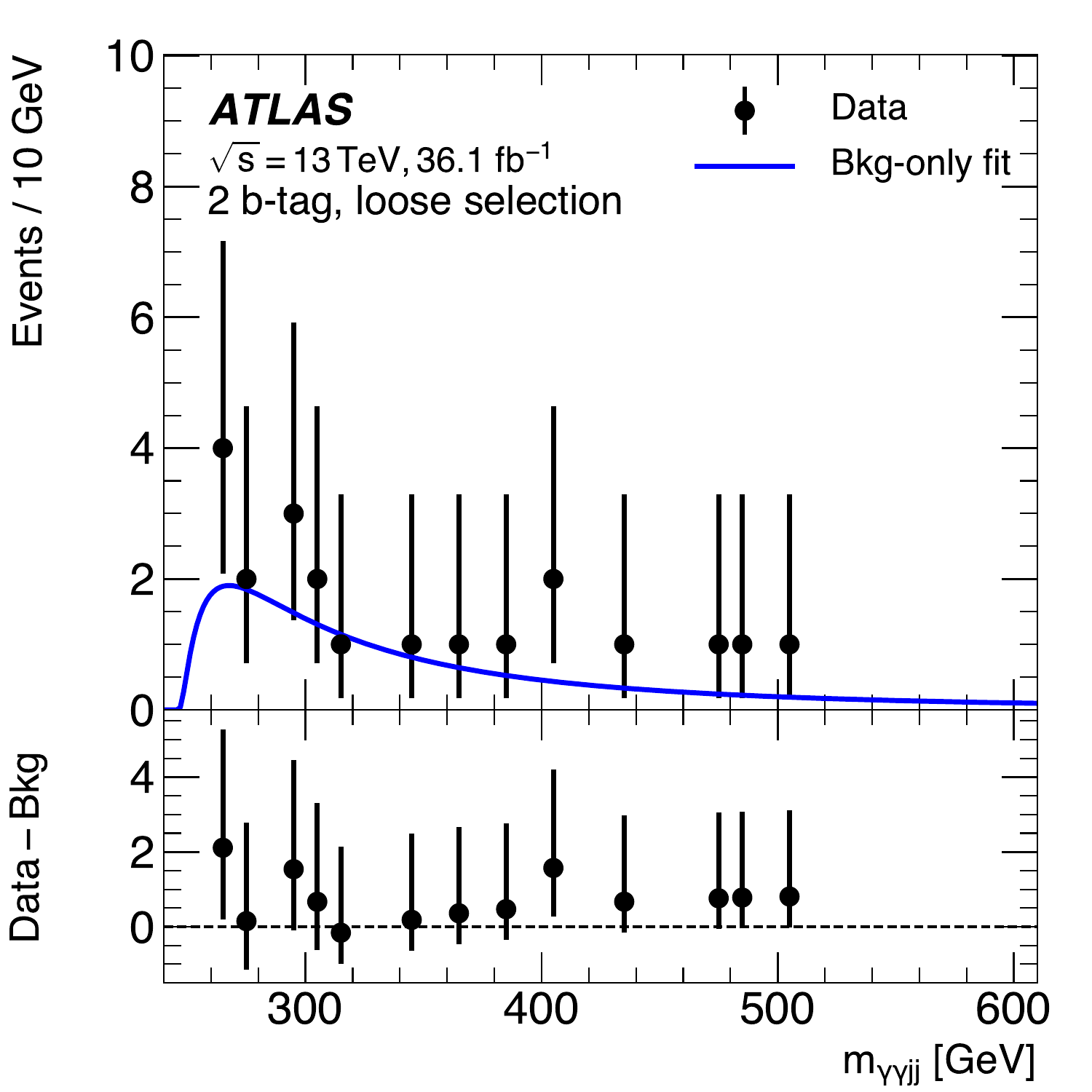}
\caption{\label{fig:bbyy-third} Distribution of $m_{\gamma\gamma jj}$ in signal regions of the search for 
$HH \to bb\gamma\gamma$ in ATLAS, with one (left) or two (right) $b$-tags, after applying a loose event selection~\cite{Aaboud:2018ftw}.}
\end{figure}

\begin{figure}[htbp]
\centering
\includegraphics[height=0.25\textheight]{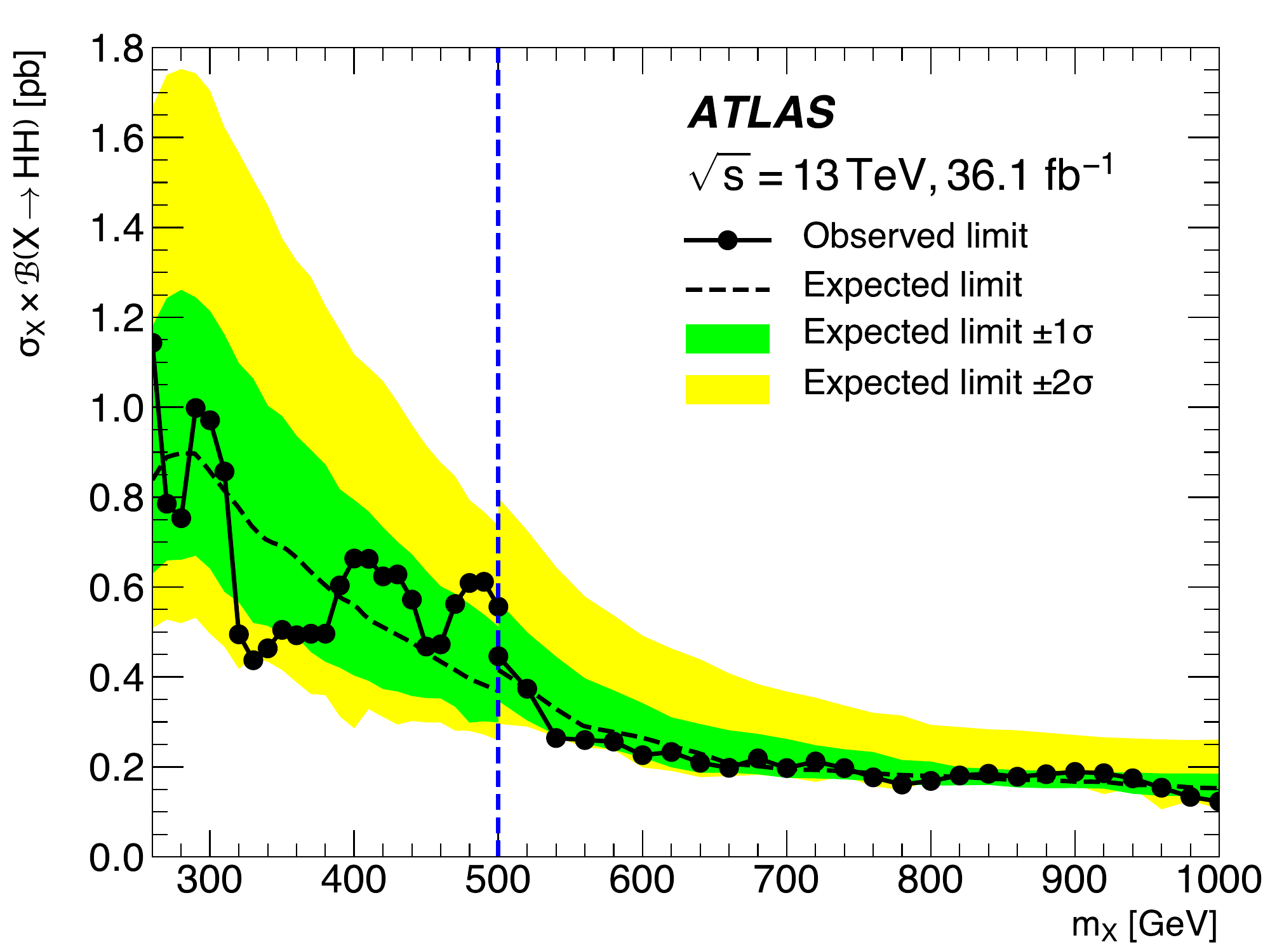}
\caption{\label{fig:bbyy-fourth} 95\% CL limits on Higgs boson pair production from a spin-0 resonance, as a function of the hypothetised 
mass, from the search for $HH \to bb\gamma\gamma$ in ATLAS. The vertical blue dashed line indicates a change in event 
selections~\cite{Aaboud:2018ftw}.}
\end{figure}

The analysis strategy used in the search for resonant $HH \to bb\gamma\gamma$ production is to extract the signal from 
the $m_{\gamma\gamma jj}$ distribution, after scaling the di-jet four-momentum by $m_{H}/m_{jj}$. The resonant $HH$ signal 
is modelled with a Gaussian distribution with exponential tails. The single-$H$ and SM non-resonant $HH$ backgrounds are taken 
from simulation. On the other hand, the continuum background of multi-jet and multi-photon events is modelled by a fit to the data, 
with a fit function chosen to minimise the spurious signal. Figure~\ref{fig:bbyy-third} shows the $m_{\gamma\gamma jj}$ distribution 
in the signal region with one and two $b$-tags, obtained with the loose selection. In the absence of a statistically significant excess 
with respect to the SM prediction, 95\% CL exclusion limits are set of the resonant Higgs boson pair production cross-section, see 
Figure~\ref{fig:bbyy-fourth}.

\section{Search for $HH \to WW\gamma\gamma$ in ATLAS}

The good mass resolution of the di-photon system arising from a $H \to \gamma\gamma$ decay has also been exploited to 
search for $HH \to WW\gamma\gamma$ in ATLAS, where one $W$-boson decays hadronically and the other one into a charged 
lepton (electron or muon) and a neutrino. The trigger, as well as the selection and reconstruction of the two photons coming from 
the $H \to \gamma\gamma$ decay are similar to those employed in the $HH \to bb\gamma\gamma$ search, except that 
$m_{\gamma\gamma}$ is required to be within 3.4~GeV of $m_H$ when searching for both non-resonant and resonant 
Higgs boson pair productions. In addition, a requirement $p_{\mathrm{T}}^{\gamma\gamma} > 100~\mbox{GeV}$ is 
used in the search for SM non-resonant production and in the search for a spin-0 resonance heavier than 400~GeV. 
In order to select $H \to WW$ decays, events are also requested to have at least one electron or muon with 
$p_{\mathrm{T}} > 10~\mbox{GeV}$, and at least two central jets which are not $b$-tagged. Both the $HH$ signal 
and the single-$H$ backgrounds are taken from simulation, and their $m_{\gamma\gamma}$ spectrum is parameterised 
with a double-sided Crystal-Ball function. On the other hand, the continuum background of multi-jet and multi-photon events 
is modelled by a fit to the data, with a second-order exponential function, chosen to minimise the spurious signal. \\

\begin{figure}[htbp]
\centering
\includegraphics[height=0.21\textheight]{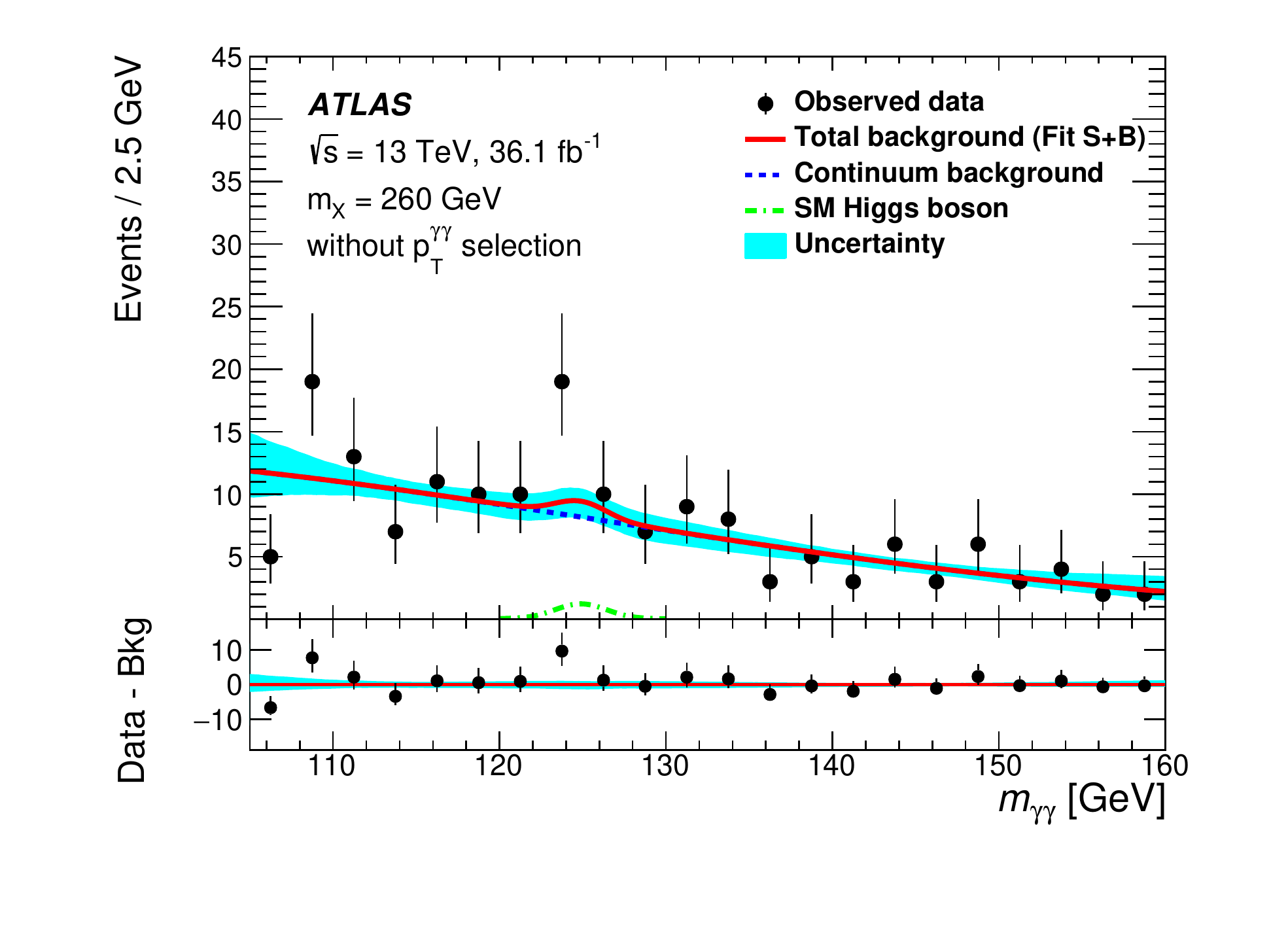}
\includegraphics[height=0.21\textheight]{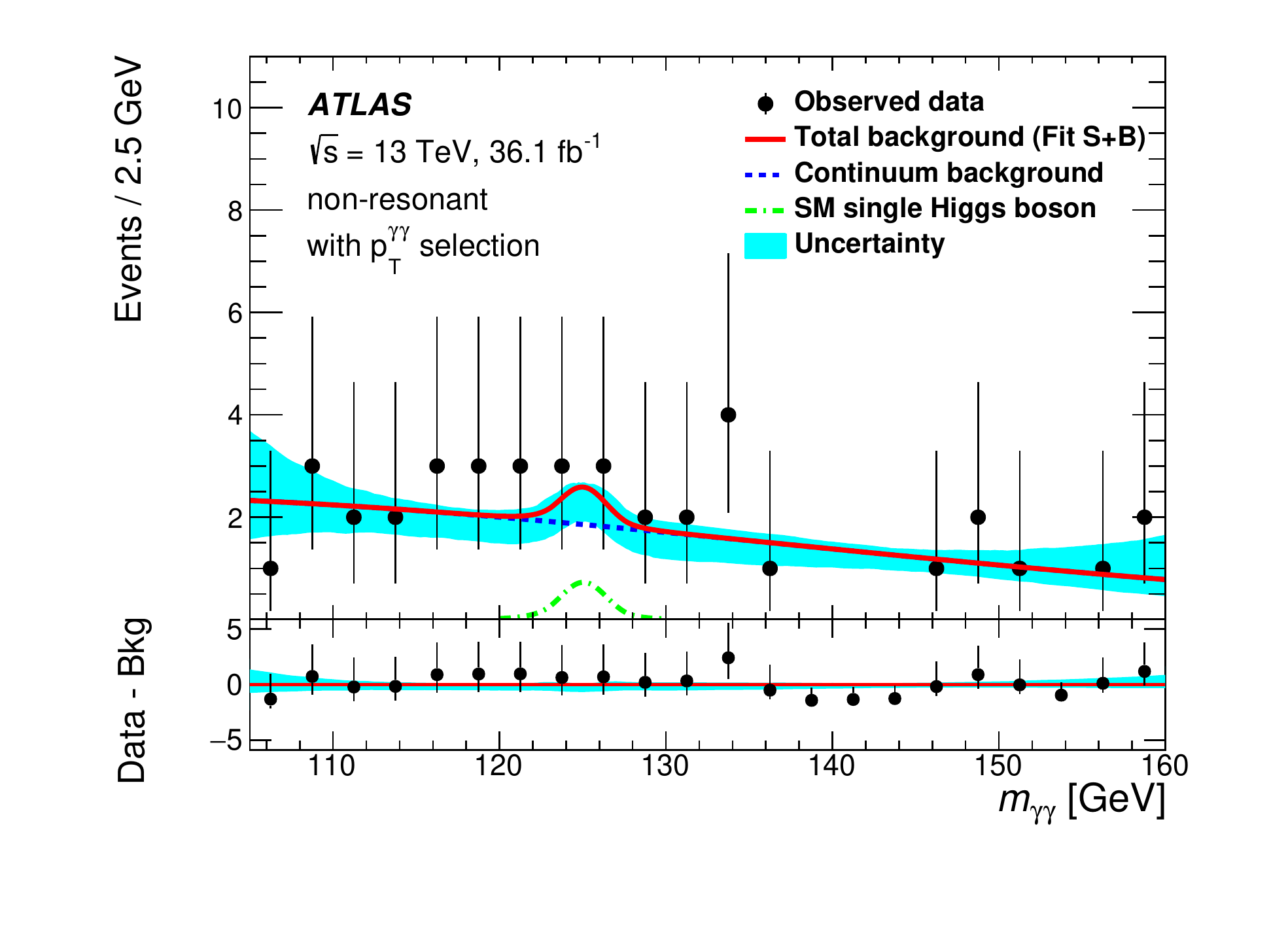}
\caption{\label{fig:WWyy-first} Di-photon mass spectrum in signal regions of the search for  $HH \to WW\gamma\gamma$ in ATLAS, 
without (left) or with (right) a cut on $p_{\mathrm{T}}^{\gamma\gamma}$~\cite{Aaboud:2018ewm}.}
\end{figure}

\begin{figure}[htbp]
\centering
\includegraphics[height=0.24\textheight]{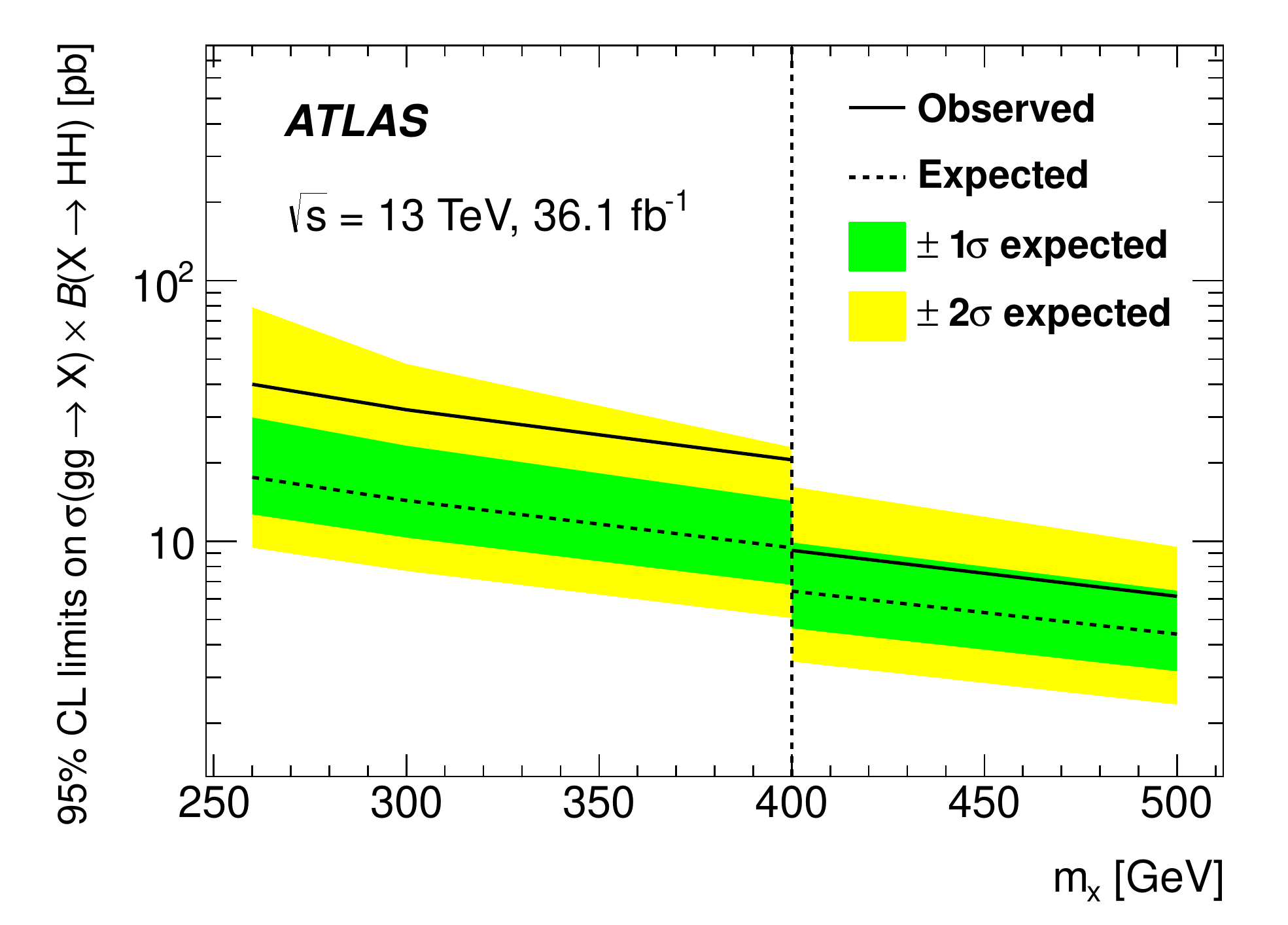}
\caption{\label{fig:WWyy-second} 95\% CL limits on Higgs boson pair production from a spin-0 resonance, as a function of the 
hypothetised mass, from the search for $HH \to WW\gamma\gamma$ in ATLAS. The vertical dashed line indicates a change 
in event selections~\cite{Aaboud:2018ewm}.}
\end{figure}

Figure~\ref{fig:WWyy-first} shows the predicted and measured $m_{\gamma\gamma}$ distribution when using or not the cut 
$p_{\mathrm{T}}^{\gamma\gamma} > 100~\mbox{GeV}$. In the absence of a statistically significant excess with respect to the 
SM prediction, 95\% CL exclusion limits are set of the non-resonant Higgs boson pair production cross-section, in units of 
the SM prediction (see Table~\ref{tab:WWyy}), and on the cross-section for a spin-0 resonance decaying to $HH$ (see 
Figure~\ref{fig:WWyy-second}).

\begin{table}[htbp]
\begin{center}
\caption{95\% CL limits on non-resonant Higgs boson pair production, from the search for $HH \to WW\gamma\gamma$ in ATLAS.}
\vspace*{2mm}
\begin{tabular}{cccccc}
\hline
Observed & $-2\sigma$ & $-1\sigma$ & Expected & $+1\sigma$ & $+2\sigma$ \\
\hline
230 & 90 & 120 & 160 & 240 & 340 \\
\hline
\end{tabular}
\label{tab:WWyy}
\end{center}
\end{table}

\section{Conclusion}

These proceedings report on three searches for Higgs boson pair production in up to 36.1~fb$^{-1}$ of LHC $pp$ collision 
data, which have been recently published by the ATLAS Collaboration: $HH \to bbbb$, $HH \to bb\gamma\gamma$ and 
$HH \to WW\gamma\gamma$. The data were analysed to search for non-resonant production of $HH$ pairs as well as 
for heavy resonances decaying to two SM-like Higgs bosons, with various analysis strategies and event selections. The result 
of a fourth search, based on the $HH \to bb\tau\tau$ final state, was published after the conference~\cite{Aaboud:2018sfw}, 
reaching 95\% CL exclusion limits of 12.7 times the SM expectation for the non-resonant $HH$ production mode. A statistical 
combination of the three most sensitive searches was also performed by ATLAS recently~\cite{ATLAS-comb}. The combined 
observed limit on the non-resonant Higgs boson pair cross-section is 0.223~pb at 95\% CL, which is equivalent to 6.7 times the 
predicted SM cross-section. The ratio $\kappa_{\lambda}$ of the Higgs boson self-coupling to its SM expectation is constrained 
at 95\% CL to $-5.0 < \kappa_{\lambda} < 12.1$. The search for Higgs boson pair production will remain at the core of the ATLAS 
research program towards the end of the LHC Run-2 and beyond, as it allows to probe directly the Higgs potential as well as 
search for new physics in the Higgs sector.\\

Copyright [2018] CERN for the benefit of the ATLAS Collaboration. Reproduction of this article or parts of it is
allowed as specified in the CC-BY-4.0 license.

\end{document}